# Machine Learning Based Fast Power Integrity Classifier

HuaChun Zhang, *Member, IEEE*, Lynden Kagan, *Member, IEEE*, Chen Zheng, *Member, IEEE*

**Abstract**

**In this paper, we proposed a new machine learning based fast power integrity classifier that quickly flags the EM/IR hotspots. We discussed the features to extract to describe the power grid, cell power density, routing impact and controlled collapse chip connection (C4) bumps, etc. The continuous and discontinuous cases are identified and treated using different machine learning models. Nearest neighbors, random forest and neural network models are compared to select the best performance candidates. Experiments are run on open source benchmark, and result is showing promising prediction accuracy.**

## 1. Introduction

In modern integrated circuit design, due to the ever requirement for higher performance, power density keeps going up. Thus, larger currents, narrower wires all lead to drastic increase in IR drop and electromigration (EM) reliability degradation. To ensure chip work functionally, both IR drop and current density need to be below certain threshold (e.g. 10% of supply voltage for IR threshold). Sign-off tools are usually run at the end to check full-chip power integrity, including IR and EM [1]. IR drop violation will result in transistor switch failure [2], while EM violations will cause open or short in interconnects [3]. However, this will significantly increase turn around time as power integrity sign-off check is highly time consuming, and for any IR or EM violation, designers need to perform engineering-change-order (ECO) to fix the violations or even go back to place or floorplan stage to fix all violations [4]. Such iteration may need to loop for multiple times to eventually achieve power integrity clean design [5-6]. This extends turn around time severely and is extremely detrimental to design cycle. Therefore, a fast power integrity check across different stages is desired to help designers detect potential IR/EM issue early and reduce the effort for closing chip power integrity.

## 2. Motivation

There have been many research works going on for power grid analysis. In [7], an incremental power grid analysis is studied; in [8] a power grid method for flip-chip design is provided; in [9], a macro circuit model is provided to help power grid analysis. However, all previous works focused on how to utilize numerical computation technique to speed up analytical calculation with high accuracy, thus are more sign-off oriented approaches. They mainly reduce runtime during sign-off check, but do not change the design sequence, therefore the benefits of turn around time are minor. In our work, we try to find a solution that can be applied to any design stage after placement, and can provide some level of power integrity confidence that allow designers to quickly identify potential EM/IR hotspots and provide fixes at early stages. According to the target and needs, we make a few assumptions as prerequisites to investigate further our fast power integrity analysis.

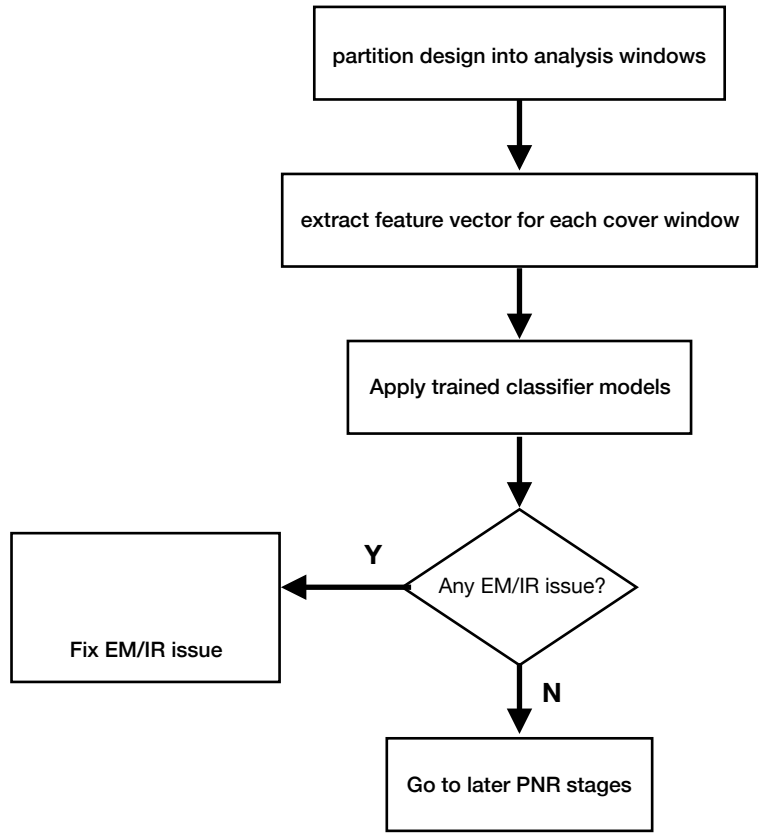

**Fig 1. Flow chart of machine learning based EM/IR analysis**

*(1) analysis results needs to be accurate, but no exact IR drop or current density value is needed, since we only care if an instance is a hotspot or not;*

*(2) power grid design is uniform, thus voltage and current distribution is continuous and within certain limit of gradient;*

*(3) as controlled collapse chip connection (C4) package is used, certain level of locality for EM/IR analysis is assumed, e.g. cell instance drop within given a 5um x 5um window can be approximately determined by its expanded cover window of 20um x 20um.*

Based on above assumptions, we found that a machine learning based binary classifier is a perfect match for the desired fast power integrity check framework [10]. Figure 1 illustrates the flow chart for the proposed methodology. In Section 3, we will discuss how to select features to describe the power grid and power density distribution, as well as the signal routing impact. In Section 4, compared performance behaviors of a few machine learning models and select the best fit in our case. In Section 5, we provided experimental results for a few open source real design blocks. In Section 6, we draw conclusions and discuss directions for future works.

## 3. Feature Selection

To completely describe the entire power delivery and consumption network, we divide it into three parts: (1) power delivery network; (2) cell power density map; (3) signal routing representation. Figure 2 illustrate the top view of a typical cell placement, power grid and signal routing within an analysis window.

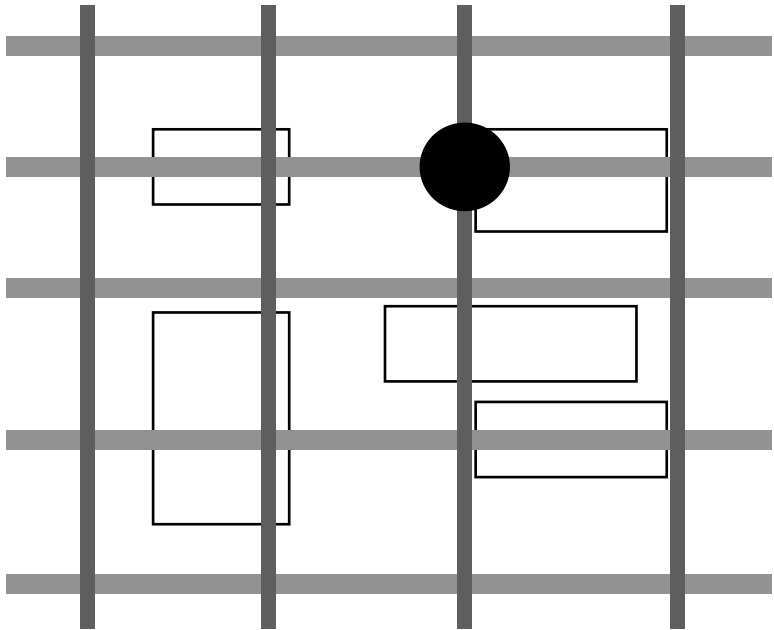

**Fig 2. Top view of power grid, cell placement and C4 bumps**

For part (1), when given the power stripe width, pitch and offset for each layer, a power delivery network can be complete described. Also, for part (2), when cell placement is determined, cell power density map can also be completed extracted without information loss. Note that here we mainly focus on static EM/IR analysis. For (3), since signal routing is not considered as first order impact to EM/IR analysis, we can compress the information for feature extraction convenience. The main contribution of signal routing to power analysis is for parasitic coupling capacitance, thus a natural feature extraction to represent the signal routing for power analysis is the total coupling capacitance within the analysis window.

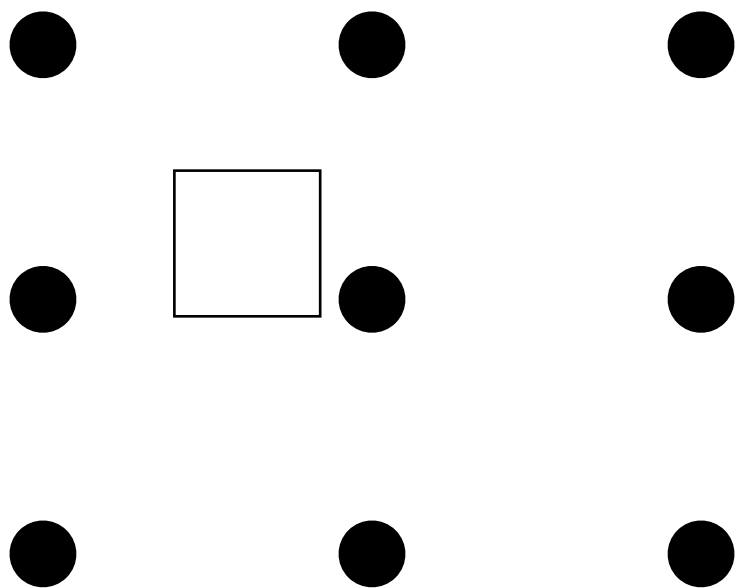

**Fig 3. Analysis window and C4 bump array**

Based on above discussion, we propose the input feature vector as described in equation (1):

$$FV = [w_i, p_i, o_i(i=1,2,...,T), pd_{i,j}(i=1,2,...,L/W; j=1,2,...,L/H), c, X_i, Y_i(i=1,2,...,9)] \quad (1)$$

where FV is the feature vector, $w_i$, $p_i$ and $o_i$ are width, pitch and offset of metal layer i, T is the index of top metal layer; $pd_{i,j}$ is the power density of a sub window i,j, L is the expanded cover window step length, W is the cell width of a unit size inverter, H is the standard cell row height; c is the total coupling capacitance of the signal routing; $X_i$, $Y_i$ is the distance from the center of the cover window to the nine nearest C4 bumps. As C4 bumps on considered ideal voltage source for on-chip power grid analysis, their locations are crucial to provide boundary conditions. Figure 3 illustrates the relation between the cover window and the C4 bumps. Moreover, as variation effect is becoming more and more significant in modern designs [11][12], we could also include variation effects into the feature vectors. Note that a cover window will always be contained within a nine C4 bump matrix [13]. Also, temperature could also be an important factor that impact chip power integrity [14], it also worth capturing these effects in the features.

## 4. Sample and Model Selection

As stated in Section 2, we made a few assumptions such that the continuity of power grid analysis is satisfied. However, there are a few exceptions, mainly the boundary case and the boundary-corner case [15]. When a cover window actually extends outside the design boundary, the continuity is lost, as the power grid is cut-off. Figure 4 illustrates the boundary and corner cases.

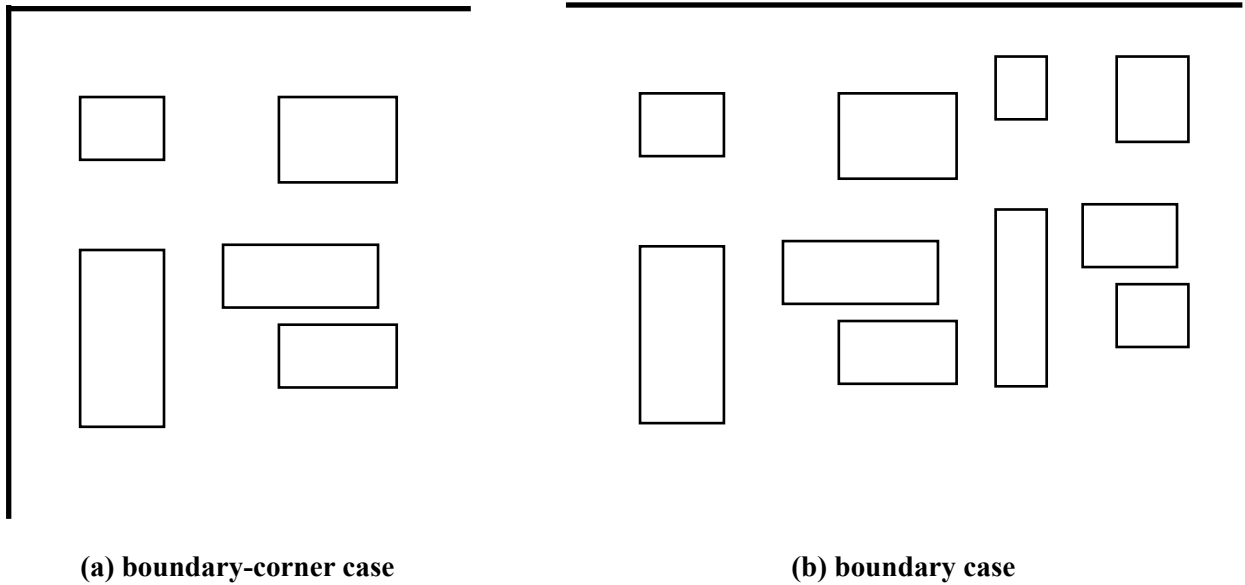

(a) boundary-corner case                              (b) boundary case

**Fig 4. Special cases for discontinuous power grid**

Equation (1) is expanded to equation (2) to describe the feature vector for boundary and corner cases:

$$FV = [X_a, Y_a, X_b, Y_b, w_i, p_i, o_i(i=1,2,...,T), pd_{i,j}(i=1,2,...,L/W; j=1,2,...,L/H), c, X_i, Y_i(i=1,2,...,9)] \quad (2)$$

where $X_a$, $Y_a$, $X_b$, $Y_b$ are lower left and upper right coordinates of the power grid bounding box referenced to the analysis window origin. These cases are trained separately as they are expected to have very different behavior as the main continuous cover windows.

We experiment with a few typical machine learning classifier models from scikit-learn package [16] with python with the extracted feature vectors. The models we explored are nearest neighbors [17], random forest [18] and neural network [19]. We also incorporate the EM-sensitive case current analysis in [20] for EM analysis.

We observed that for continuous window, nearest neighbors gives the best performance as its simplicity matches well with the continuity property. While for discontinuous window, both random forest and neural network models outperform nearest neighbors, and neural network works slightly better compared to random forest.

## 5. Experimental Results

We used one block from the open source design or1200_fpu_fcmp [21]. Physical design is implemented using Synopsis ICC2 [22]. We used RedHawk [23] for preliminary EM/IR analysis. We biased the cell power to create some different EM/IR results across the design block. Table 1 demonstrates the results using the machine learning based fast EM/IR analysis as compared to sign-off analysis. Based on those results, designers can apply various techniques to fix detected hotspots [24-27].

Table 1. Comparison between Fast EM/IR check and sign-off tool

|  | continuous window | | discontinuous window | |
| --- | --- | --- | --- | --- |
|  | sign-off | Fast check | sign-off | Fast check |
| # Analysis windows | 42560 | 42560 | 900 | 900 |
| # IR violations | 427 | 411 | 14 | 15 |
| # EM violations | 58 | 44 | 0 | 0 |
| # False positive | N/A | 13 | N/A | 3 |
| Prediction accuracy | N/A | 91.13% | N/A | 85.71% |

## 6. Conclusions

In this paper, we discuss the limitations of traditional power integrity analysis, and proposed a fast EM/IR classifier to quickly identify potential hotspot for power integrity issues. The feature selection represents the power grid and power dissipation of a given analysis window, and different models are used for different scenarios. Experimental results show good matching result between the classifier and sign-off tool. Using our proposed methodology, designers can save design effort/cycles for closing EM/IR issues and significantly reduce the design turn around time.